\documentclass[a4paper]{IEEEtran}
\usepackage{color}
\usepackage{leftidx}
\usepackage{cite}
\usepackage{microtype}
\usepackage{cleveref}
\usepackage{graphicx,amssymb,amstext,amsmath}
\usepackage[ruled,vlined,lined,ruled,linesnumbered]{algorithm2e}
\begin{document}

\title{Distributed Scheduling in Wireless Powered Communication Network: Protocol Design and Performance Analysis}

\author{Suzhi~Bi$^*$, Ying~Jun~(Angela)~Zhang$^\dagger$, and Rui~Zhang$^\ddagger$\\
$^*$College of Information Engineering, Shenzhen University, Shenzhen, Guangdong, China 518060\\
$^\dagger$Department of Information Engineering, The Chinese University of Hong Kong, Shatin, N.T., Hong Kong SAR\\
$^\ddagger$Department of Electrical and Computer Engineering, National University of Singapore, Singapore 117583\\
E-mail:~bsz@szu.edu.cn, yjzhang@ie.cuhk.edu.hk, elezhang@nus.edu.sg \vspace{-2ex}
\thanks{This work was supported in part by the National Natural Science Foundation of China (Project number 61501303), and the Foundation of Shenzhen City (Project numbers JCYJ20160307153818306). The work of Y.~J.~Zhang was supported in part by General Research Funding (Project number 14209414) from the Research Grants Council of Hong Kong and by the National Basic Research Program (973 program Program number 2013CB336701).}}

\maketitle

\begin{abstract}
Wireless powered communication network (WPCN) is a novel networking paradigm that uses radio frequency (RF) wireless energy transfer (WET) technology to power the information transmissions of wireless devices (WDs). When energy and information are transferred in the same frequency band, a major design issue is transmission scheduling to avoid interference and achieve high communication performance. Commonly used centralized scheduling methods in WPCN may result in high control signaling overhead and thus are not suitable for wireless networks constituting a large number of WDs with random locations and dynamic operations. To tackle this issue, we propose in this paper a distributed scheduling protocol for energy and information transmissions in WPCN. Specifically, we allow a WD that is about to deplete its battery to broadcast an energy request buzz (ERB), which triggers WET from its associated hybrid access point (HAP) to recharge the battery. If no ERB is sent, the WDs contend to transmit data to the HAP using the conventional $p$-persistent CSMA (carrier sensing multiple access). In particular, we propose an energy queueing model based on an energy decoupling property to derive the throughput performance. Our analysis is verified through simulations under practical network parameters, which demonstrate good throughput performance of the distributed scheduling protocol and reveal some interesting design insights that are different from conventional contention-based communication network assuming the WDs are powered with unlimited energy supplies.
\end{abstract}
\vspace{-2ex}

\IEEEpeerreviewmaketitle

\section{Introduction}
The recent development of RF-enabled WET technology provides a new solution to continuously power energy-constrained wireless devices (WDs) over the air \cite{2014:Bi,2015:Lu}. Wireless power in tens to several hundred of microwatts can be effectively transferred to WDs within ten meters distance, making self-sustainable network operation truly feasible and efficient for many low-power wireless applications, e.g., wireless sensor networks and RF identity (RFID) systems with a large number of WDs. The application of WET to wireless communications spurs a novel networking structure named wireless powered communication network (WPCN), where the WDs transmit information using the energy harvested by means of WET \cite{2016:Bi1}. WPCN removes the need of frequent battery replacement/recharging and reduces the probability of energy outage. The network lifetime can thus be largely extended and the communication performance can also be improved with more sustainable power supply.

\begin{figure}
\centering
  \begin{center}
    \includegraphics[width=0.45\textwidth]{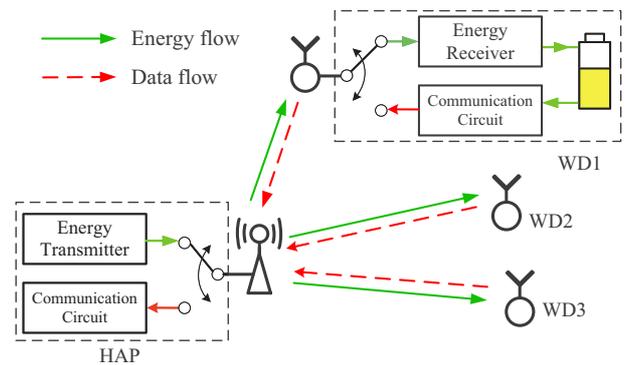}
  \end{center}
  \caption{A TDD based wireless powered communication network \cite{2014:Bi}.}
  \label{101}
\end{figure}

As shown in Fig.~\ref{101}, we consider a single-cell WPCN where a hybrid access point (HAP) is responsible for transmitting wireless energy to and receiving wireless information transmission (WIT) from a set of distributed WDs \cite{2014:Bi}. In practice, WET and WIT are desired to operate in the same frequency band to achieve higher spectrum efficiency and cost effectiveness. In this case, time-division-duplexing (TDD) circuit structures are applied at both HAP and WDs to switch between WET and WIT modes to avoid the harmful interference from WET to information decoding \cite{2013:Zhou}. While a major design challenge is transmission scheduling for WET and WIT to achieve both efficient communication and energy harvesting. Most of the existing studies in WPCN have assumed the HAP to centrally coordinate the WET and WIT with the WDs. For instance, \cite{2014:Ju1} proposes a round-robin based scheduling, where the HAP and WDs take turns to transmit energy or data. The duration of each WD's transmission is optimized by the HAP according to the global instantaneous channel state information (CSI) and then sent to all the WDs. \cite{2014:Liu2} later extends \cite{2014:Ju1} to the case with a multi-antenna HAP that enables more efficient energy beamforming technique for WET and SDMA (spatial duplexing multiple access) for WIT. A similar round-robin based scheduling method is considered in \cite{2014:Niyato}, where each energy-harvesting WD can be either active or inactive in a time slot to achieve a balance between communication delay and energy consumption. In addition, \cite{2014:Misic} considers a polling-based method that the HAP periodically inquires the WDs about their residual energy levels and performs WET whenever some WDs are in low battery state.

In practice, the above centralized methods often incur considerable signaling overhead on channel estimation, control, synchronization, etc. This could be costly in networks with a large number of WDs (e.g., sensors) that are randomly deployed and switch on/off over time for energy saving. In this case, distributed scheduling of WET and WIT is of high practical interests. Although distributed wireless charging control and data transmissions have been well investigated separately (e.g., \cite{2016:Bi,2000:Bianchi,2005:Bianchi,2013:Bi}), there are only few studies integrating them in the design of WPCN. For instance, \cite{2011:Kim} proposes an energy-adaptive CSMA-type MAC (medium access control) method, where the access probability of a WD decreases with its energy harvesting rate. However, it assumes that WET is independent of WIT, and thus no joint WET and WIT scheduling is considered. \cite{2014:Naderi} proposes a RF-MAC scheme that multiple HAPs are divided into groups to perform WET in respond to WDs' energy request, and the WDs use CSMA-type random access control to coordinate the data exchange among each other. The RF-MAC method, however, requires the WDs to bear complicated computation and channel estimation tasks. Further, \cite{2015:Tamilarasi} considers a simplified version of RF-MAC, where the throughput performance of a WPCN using a single HAP is evaluated via simulations. Nonetheless, the analysis of both works is limited and does not capture the important coupling between energy and information transmissions.

In this paper, we present a practical distributed scheduling protocol for WPCN. Similar to the idea of RF-MAC, we allow each WD that is about to deplete its battery to broadcast an energy request buzz (ERB) signal in order to trigger the WET by the HAP to recharge its battery. If no ERB is sent, the WDs then contend to transmit data to the HAP based on the conventional $p$-persistent CSMA.\footnote{$p$-persistent CSMA achieves similar performance as the exponential backoff scheme in \cite{2000:Bianchi} when the transmit probability $p_t$ of the WDs is proportional to the user number \cite{2013:Bi1}. In practice, the HAP is aware of the number of associating WDs and thus can calculate $p_t$ and broadcast its value to the WDs.} In particular, we propose an energy queueing model to analyze the throughput performance of the proposed distributed scheduling protocol. Simulation results are provided to verify our analysis and show that the proposed method can achieve good throughput performance as compared to a benchmark $p$-persistent CSMA network assuming always sufficient energy supply. In addition, an interesting \emph{energy decoupling} property is revealed, which is useful in deriving the throughput and understanding the insight on designing distributed scheduling in WPCN.

\section{System Model}
As shown in Fig.~1, we consider a WPCN consisting of a HAP and $N$ WDs, where all the devices each have one single antenna. We assume that WET and WIT are performed over the same frequency band, such that each WD's antenna is used for both energy harvesting and communication in a TDD manner (see WD1 in Fig.~\ref{101}). The energy harvesting circuit converts the received RF signal to DC energy and stores in a rechargeable battery. The energy is then used to power the WIT. The HAP also has a similar TDD circuit structure (see Fig.~\ref{101}) to switch between energy transfer and communications with the WDs.

We assume that all the $N$ WDs are continuously backlogged, i.e., they always have packets to transmit. Besides, the network is fully connected, such that the transmission of one device (WD or HAP) can be overheard by all the other devices. Meanwhile, all devices are assumed to have carrier sensing capability, such that they remain silent when sensing any ongoing energy/information transmission and attempt to transmit only after the channel becomes idle. The proposed distributed scheduling mechanism is illustrated in Fig.~$2$ and explained as follows.

\begin{figure}
\centering
  \begin{center}
    \includegraphics[width=0.45\textwidth]{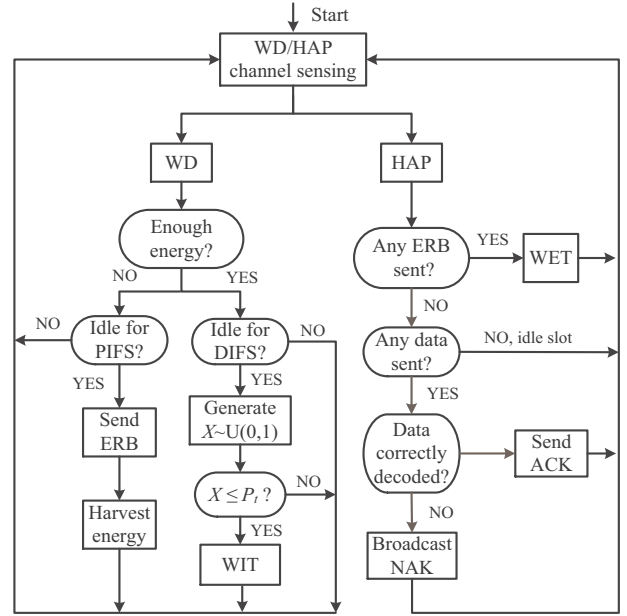}
  \end{center}
  \caption{A diagram illustrating the distributed information and energy scheduling in WPCN.}
  \label{102}
\end{figure}

\subsection{Distributed WET and WIT Scheduling}
In the proposed distributed scheduling mechanism, each WD continuously monitors its residual battery level. If it is above a predetermined threshold, the WD waits for the channel to be continuously idle for a DIFS (distributed inter-frame spacing) time and then transmits independently a payload packet with probability $0<p_t<1$ to the HAP. The duration of DIFS is much larger than the signal round-time-delay (RTD) of the network, such that a WD's data transmission will not interfere with the potential data transmissions of the other WDs due to signal prorogation delay. The packet header contains the identity of the transmitting WD, such that the HAP can identify the sender if the packet is successfully decoded. Otherwise, if a WD finds its residual battery level below the threshold, it waits the channel to be continuously idle for a PIFS (priority inter-frame spacing) time and then sends a short energy request buzz (ERB) signal. The duration of the PIFS is smaller than that of DIFS (but still much larger than the RTD), such that a higher priority is assigned to sending an ERB than a data payload.

From the HAP's perspective, it can identify the current time slot as a WET slot when sensing any ERB sent, either by one or multiple WDs, and respond by performing WET for $T_{et}$ amount of time. Meanwhile, all the WDs sensing the ERB signal switch to energy harvesting mode. Otherwise, if no ERB signal is sensed, the HAP identifies the current time slot as a WIT slot and switches to information receiving mode. Due to the close communication range (say, within $10$ meters) typically for WPCN, we assume the receiver signal-to-noise ratio (SNR) is sufficiently high and thus neglect the decoding errors caused by channel fading and receiver noise during data transmissions. Accordingly, a WIT slot may correspond to one of the following three scenarios:
\begin{enumerate}
  \item success: if only one WD transmits data, the transmitted packet can be successfully decoded by the HAP, which then responds to the transmitting WD by sending an ACK message (containing the ID of the transmitter) after sensing the channel to be idle for SIFS (short inter-frame spacing) time. Notice that SIFS is shorter than the DIFS and PIFS such that no WD will transmit before the HAP sends the ACK;
  \item collision: if more than one WDs transmit data in the same time slot, the multiple packets will collide and none of then can be correctly decoded by the HAP. In this case, the HAP broadcasts a NAK message after sensing the channel to be idle for SIFS. By doing so, the transmitting WDs can identify collision and schedule data retransmission;
  \item idle: otherwise, no WD transmits data and all WDs keep silent for a mini slot of duration $\sigma$. Notice that the WDs do not need to wait for another DIFS time to transmit data after an idle time slot. Instead, they can persistently access the channel with probability $p_t$ in the following mini slots until some WD transmits. Therefore, we may observe consecutive idle mini slots.
\end{enumerate}
After the current time slot, either for WET or WIT, all the devices continue to sense the wireless medium, and the above iteration repeats itself.

We use an example to illustrate the operation of the proposed protocol in Fig.~3. Initially, the $2$ WDs have sufficient energy to transmit and WD1 transmits successfully. After the channel becomes idle for DIFS, no WD transmits in two consecutive idle time slots each of duration $\sigma$, until they both transmit data in the third attempt and cause a collision. Then, after transmitting data, WD$1$ is lack of energy and sends an ERB signal after a PIFS. Upon detecting the ERB signal, the HAP starts WET for the WDs to harvest energy.

\subsection{Wireless Energy and Information Transfer Model}
For a WET time slot, we assume channel fading effect is averaged out over the duration of energy transmission such that the received energy by the $n$-th WD is only related to its distance $d_n$ to the HAP. Due to the broadcasting nature of wireless channels, all the $N$ WDs can harvest energy in a WET time slot. We also assume that the WDs cannot harvest energy from WIT, as transmit power of WD is significantly lower than that of WET by the HAP, e.g., $10$mW versus $3$W. Thus, the received energy by the $n$-th WD in a WET slot is
\begin{equation}
\label{24}
R_n = \eta A_d P_h T_{et} \left(\frac{3\cdot 10^8}{4\pi f_d d_n}\right)^\upsilon , \ n=1,\cdots,N,
\end{equation}
where $\eta\in(0,1)$ denotes the energy harvesting efficiency, $A_d$ denotes the antenna gain, $P_h$ denotes the power of WET, $f_d$ denotes the carrier frequency and $\upsilon\geq 2$ denotes the path loss exponent, which is assumed equal for all the WDs.

For a WIT slot, we assume that all the WDs transmit with constant power $P_w$. For the simplicity of analysis, we assume that each WD transmits a payload of fixed duration $T_{pl}$ in a WIT slot, such that it consumes $V_n = P_w T_{pl}$ amount of energy regardless of that the transmission is successful or results in a collision. Nonetheless, our analysis can also be extended to the case that the payload lengths are different in a success and a collision slot, such as a CTS/RTS-like scheme in 802.11 WLAN \cite{2000:Bianchi}. With channel error neglected, the decoding failure is only caused by transmit collisions.

\begin{figure}
\centering
  \begin{center}
    \includegraphics[width=0.5\textwidth]{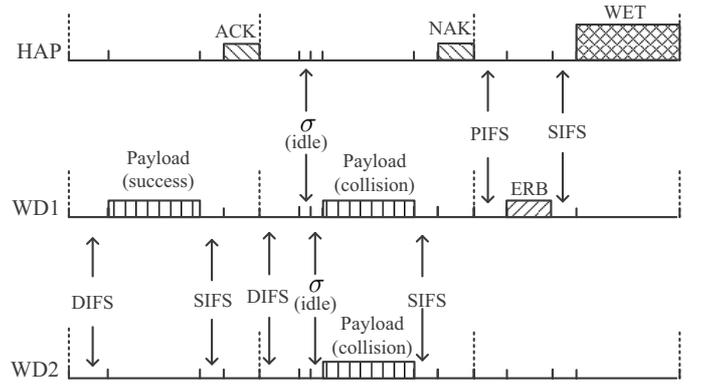}
  \end{center}
  \caption{A $3$-node example of the distributed scheduling protocol operation.}
  \label{103}
\end{figure}

\subsection{Device Battery Model}
In this paper, we consider a discrete energy model and assume that the transmission of each fixed-length payload consumes $1$ unit of energy and the battery capacity of each WD is $C$ units, where $C>>1$ is a positive integer. Besides, we assume that the $n$-th WD harvests fixed $e_n$ units of energy in each WET slot.\footnote{Here it means the energy harvested minus that spent on sending ERB signal. The energy consumption on channel sensing is also neglected for simplicity.} Here, $e_n<<C$ is a positive integer, $n=1,\cdots,N$, depending on the distance between the WD and the HAP. This may correspond to a practical design requirement that $e_n\geq 1$, $\forall n$, to avoid frequent energy transmissions, which is achievable through either setting a long enough $T_{et}$ or preventing ineffective far-away WDs from associating with the HAP. We denote $B_n^l$ as the battery level (in units) of the $n$-th WD at the end of the $l$-th time slot, and $E_n^l$ and $Q_n^l$ as the number of units of energy harvested and consumed during the $l$-th time slot, respectively. Then, the battery dynamics of the $n$-th WD can be expressed as
\begin{equation}
B_n^{l} = \min\left\{\max\left(B_n^{l-1}+E_n^{l} - Q_n^{l},0\right),C\right\},
\end{equation}
where $l=1,2,\cdots$ and $B_n^{0}$ denotes the initial energy level. Depending on the type of the $l$-th transmission slot, $E_n^{l}$ and $Q_n^{l}$ can be categorized as follows:
\begin{enumerate}
  \item WET slot: $E_n^{l}=e_n$ and $Q_n^{l} =0$ for all the WDs;;
  \item Success/collision WIT slot: $E_n^{l}=0$ and $Q_n^{l} = 1$ for transmitting WDs, and $E_n^{l}= Q_n^{l} = 0$ otherwise;
  \item Idle WIT slot: $E_n^{l}= Q_n^{l} = 0$ for all the WDs.
\end{enumerate}
Without loss of generality, we assume that a WD sends ERB signal when $B_n^{l} =0$.

\subsection{Performance Metric}
In this paper, a key performance metric is the normalized network throughput, defined as the percentage of air time occupied by successful data transmissions expressed as
\begin{equation}
\label{1}
\psi= \frac{P_{suc} \cdot T_{suc}}{P_{suc}T_{suc}+ P_{col}T_{col} + P_{idl}T_{idl} + P_{ene}T_{ene}},
\end{equation}
where $\{P_{suc}, P_{col}, P_{idl}, P_{ene}\}$ and $\{T_{suc}, T_{col}, T_{idl}, T_{ene}\}$ denote the probabilities of a successful packet transmission slot, a packet collision slot, an idle slot, and an energy transfer slot, respectively. By assuming the durations of ACK and NAK are equal, we can see from Fig.~3 that
\begin{equation*}
\begin{aligned}
&T_{suc} = T_{col} = DIFS + T_{pl} + SIFS + ACK,\\
&T_{idl} = \sigma, T_{ene} = PIFS + ERB + SIFS+ T_{et}.
\end{aligned}
\end{equation*}
In the next section, we analyze the throughput performance of the distributed energy and information scheduling algorithm.

\section{Throughput Performance Analysis}

\subsection{Energy Queueing Model}
We start with modeling the battery dynamic of each WD as a B-D queueing process. As shown in Fig.~4, we drop the superscript $l$ for simplicity of expression and use $B_n$ to denote the residual energy of the $n$-th WD at the beginning of a time slot. We refer to the WD as in the $i$-th energy state if $B_n=i$, $i \in \left\{0,1,\cdots, C\right\}$. In particular, we use $p_n^e(i)$ to denote the probability of the $n$-th WD in the $i$-th energy state observing a WET slot, $i=1,\cdots,C$. Therefore, we can express the state transition probability $p_n(i\rightarrow j)$, which denotes the probability that the $n$-th WD changes from the $i$-th to the $j$-th energy state, as follows:
\begin{subequations}
\label{7}
\begin{align}
&p_n(i\rightarrow \max\{i+e_n,C\}) = p_n^e(i),\label{4}\\
&p_n(i\rightarrow i-1) = p_t\left(1-p_n^e(i)\right),\\
&p_n(i\rightarrow i)= \left(1-p_t\right)\left(1-p_n^e(i)\right)\label{5},
\end{align}
\end{subequations}
for intermediate states with $0<i<C$. Besides, the other two boundary states $0$ and $C$ satisfy
\begin{subequations}
\label{8}
\begin{align}
&p_n(0\rightarrow e_n) = 1, \label{2}\\
&p_n(C\rightarrow C-1) = p_t\left(1-p_n^e(i)\right),\label{6}\\
&p_n(C\rightarrow C)= 1- p_t\left(1-p_n^e(i)\right) \label{3}.
\end{align}
\end{subequations}
Here, (\ref{4})-(\ref{5}) correspond to a WET slot, a success/collision WIT slot, and an idle slot, respectively. In addition, (\ref{2}) holds because a WD with $B_n=0$ will immediately send ERB and receive energy in the current time slot. (\ref{6}) and (\ref{3}) hold because the energy of a fully-charged WD will reduce only when it transmits data, and remain unchanged otherwise.

\begin{figure}
\centering
  \begin{center}
    \includegraphics[width=0.5\textwidth]{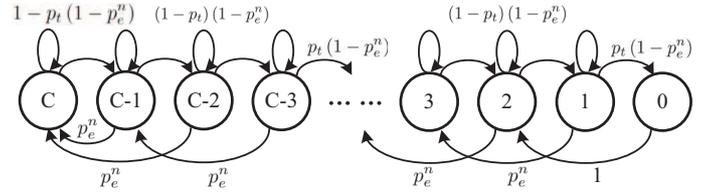}
  \end{center}
  \caption{Energy queueing model of the $n$-th WD ($e_n=2$).}
  \label{104}
\end{figure}

It is evident that the occurrence of an energy transfer slot is related to the energy states of all the WDs, where a WET slot occurs when $B_n =0$ for some WD $n$. Accordingly, precise system-level analysis requires a high-dimensional Markov system that jointly considers the energy states of all the WDs. This, however, renders the problem analytically intractable due to the large number of inter-connected states. For tractable analysis, we make in this paper the following \emph{energy decoupling assumption}.

\textbf{Energy decoupling assumption (EDA):} \emph{In the considered energy queueing system (\ref{7}) and (\ref{8}), the limiting probabilities of the $N$ WDs are independent and each WD $n$ observes a constant probability of WET in a time slot independent of its current energy state, i.e., $p_n^e(i)=p_n^e$, $i=1,\cdots,C$.}

\textbf{Remark 1}: The EDA assumption considered in this paper is analogous to the well-known \emph{mean-field decoupling assumption} made in the seminal work on performance analysis of 802.11 DCF medium access control \cite{2000:Bianchi}, where WDs with unlimited energy supply transmit data following a random backoff mechanism. Specifically, it assumes that when the number of WDs in a 802.11 network is large enough, each WD observes a constant collision probability upon transmission, which is independent of (but in fact related to) the current backoff stages of itself and the other WDs.

As an initial attempt to investigate the performance of distributed scheduling of WET and WIT in WPCN, we leave the proof of the above EDA assumption in our future work. For the time being, the EDA is verified using simulations later in Section V, where we show that this assumption approximately holds when the number of WDs is not too small, e.g., $N\geq 6$.

\subsection{Queueing Analysis}
With the EDA assumption, we can replace $p_n^e(i)$'s with $p_n^e$ in (\ref{7}) and (\ref{8}). We denote the steady-state limiting probabilities of the $n$-th WD as $w_n^i$, $i=0,\cdots,C$. For such a birth-death (B-D) queueing process in Fig.~4, its limiting probabilities satisfy the following equalities by establishing ``flow conservation" conditions between two adjacent states, i.e.,
\begin{subequations}
\label{9}
\begin{align}
& p_t\left(1-p_n^e\right) w_n^1 = w_n^0, \label{14}\\
& p_t\left(1-p_n^e\right)  w_n^i = w_n^0 + p_n^e\mathsmaller\sum_{j=1}^{i-1} w_n^{j},\ i= 2,\cdots,e_n, \label{15}\\
&  p_t\left(1-p_n^e\right)  w_n^i  = p_n^e  \mathsmaller\sum_{j=1}^{e_n} w_n^{i-j},\ i= e_n+1, \label{16}\cdots,C.
\end{align}
\end{subequations}
The above $C$ equations, combined with the total probability condition $\mathsmaller\sum_{i=0}^N w_n^i = 1$, can be expressed as
\begin{equation}
\mathbf{H}_n\mathbf{w}_n = \mathbf{b},
\end{equation}
where $\mathbf{w}_n = \left(w_n^0,\cdots,w_n^C\right)^T$, $\mathbf{b} = \left(0,\cdots,0, 1\right)^T$, with $(\cdot)^T$ denoting the matrix transpose and
\begin{equation*}
\label{10}
\mathbf{H}_n=\left(
  \begin{array}{ccccccc}
    1               & -\alpha_n          & 0          & 0            & 0         &   \cdots     & 0 \\
    1               & p_e^n              & -\alpha_n  & 0            & 0         &   \cdots     & 0\\
    0               & p_e^n              & p_e^n      & -\alpha_n    & 0         &   \cdots     & 0 \\
    0               & 0                  & p_e^n      & p_e^n        & -\alpha_n &   \cdots     & 0 \\
    \vdots          & \ddots             & \ddots     & \ddots       & \ddots    &   \ddots     & \vdots \\
    0               &         0          & \cdots     &        0     & p_e^n &    p_e^n         & -\alpha_n\\
    1               &         1          & \cdots     &         1    & 1         &           1  & 1 \\
  \end{array}
\right).
\end{equation*}
Here, $\alpha_n \triangleq p_t\left(1-p_n^e\right)$. Because $\mathbf{H}_n$ is a full-rank square matrix, we can obtain the steady state limiting probabilities as $\mathbf{w}_n = \mathbf{H}_n^{-1}\mathbf{b}$, with $(\cdot)^{-1}$ denoting the matrix inverse. In particular, we can infer that
\begin{equation}
\label{12}
w_n^0 = \left[\mathbf{H}_n^{-1}\right]_{1,C+1},
\end{equation}
where $[\cdot]_{i,j}$ denotes the $(i,j)$-th entry of a matrix.

Notice that the value of $\mathbf{H}_n$ is determined by $p_n^e$ and $e_n$ for the $n$-th WD. Therefore, when $e_n$ is a fixed parameter, we can expressed $w_0$ in (\ref{12}) as a function of $p_n^e$, denoted by
\begin{equation}
\label{13}
w_n^0 = f_n(p_n^e), \ n=1,\cdots,N.
\end{equation}
In general, $f_n(p_n^e)$ is a polynomial function of $p_n^e$. For instance, when $e_n=2$ and $C=3$, $f_n(p_n^e)$ can be expressed as
\begin{equation*}
\begin{aligned}
\frac{p_t^3\left(1-p_n^e\right)^3}{p_t^3\left(1-p_n^e\right)^3+ 2p_t^2\left(1-p_n^e\right)^2 + 3p_tp_n^e\left(1-p_n^e\right)+\left(p_n^e\right)^2}.
\end{aligned}
\end{equation*}
Besides, we show in the appendix that $f_n(x)$ is a decreasing function for $x\in(0,1)$.

\subsection{Throughput Derivation}
Notice that each WD $n$ with $B_n>0$ observes an ongoing ERB signal when at least one of the other $(N-1)$ WDs is in the $0$-th energy state. Accordingly, we can express $p_n^e$ as
\begin{equation}
\label{18}
p_n^e = 1 - \mathsmaller\prod_{i\neq n } \left(1-w_i^0\right) \triangleq g_n(\mathbf{w}^0), \ n=1,\cdots,N,
\end{equation}
where $\mathbf{w}^0 = \left[w_1^0,\cdots,w_N^0\right]^T$. (\ref{18}) implies that $g_n$ is a non-decreasing function of each entry in $\mathbf{w}^0$. By stacking the $N$ equations in (\ref{13}) and $N$ equations in (\ref{18}), we have
\begin{equation}
\label{19}
\mathbf{w}^0 = \mathbf{f}\left(\mathbf{g}\left(\mathbf{w}^0\right)\right) \triangleq \Psi(\mathbf{w}^0),
\end{equation}
where $\mathbf{g}(\mathbf{w}^0) = [g_1(\mathbf{w}^0),\cdots,g_N(\mathbf{w}^0)]^T$ and $\mathbf{f}(\mathbf{x}) = [f_1(x_1),\cdots,f_N(x_N)]^T$. Evidently, $\Psi$ is a non-increasing function of $\mathbf{w}^0 \in (0,1)^N$ due to the monotonic property of $f_n$. For instance, when the WDs are homogeneous, i.e., $e_n$'s are equal for all the WDs, we can denote by symmetry that $w^0 \triangleq w_n^0$ and $p^e \triangleq p^e_n = 1 - \left(1-w^0\right)^{N-1}$, $\forall n$. In this case, as $\Psi(w^0)$ is a non-decreasing function, $w^0$ can be obtained using simple bi-section search over $w_0\in(0,1)$ until $w^0 = \Psi(w^0)$ is satisfied within a given precision level. In general, $w^0_n$'s can be obtained numerically, e.g., using the quasi-Newton method.

Given $\mathbf{w}^0$, we are ready to derive the throughput performance. Specifically, the probability of a WET slot is
\begin{equation}
\label{21}
P_{ene} = 1 - \mathsmaller\prod_{i=1}^N \left(1-w_i^0\right),
\end{equation}
i.e., at least one of the WDs sends ERB. Accordingly, the probability of an information transmission slot is $P_{it}=1-P_{ene}$. Then, the probability of a successful transmission is
\begin{equation}
\label{22}
P_{suc} = P_{it}N p_t\left(1-p_t\right)^{N-1},
\end{equation}
i.e., exactly one WD transmits information. Besides, the probabilities of an idle slot and a collision slot are respectively
\begin{equation}
\label{23}
\begin{aligned}
&P_{idl} = P_{it} \left(1-p_t\right)^{N},\\
&P_{col} = P_{it} - P_{suc} -  P_{idl}.
\end{aligned}
\end{equation}
By substituting (\ref{21})-(\ref{23}) into (\ref{1}), we can obtain the throughput $\psi$. We notice that each WD has the equal probability to transmit information in a WIT slot. Therefore, the $N$ WDs have the same average data rate $\psi/N$. From (\ref{21}), if some WD $n$ has very high probability of energy outage, i.e., large $w_n^0$, the data rates of all the WDs can be very low. Therefore, our proposed method should be applied to a network with limited WET range, e.g., the maximum WD-to-HAP distance is less than $10$ meters to ensure that all WDs can be effectively charged by the HAP.

\begin{table}
\caption{Simulation Parameters}
\footnotesize
\begin{center}
\begin{tabular}{|c|c||c|c|}
 \hline
  HAP power &   $3$ W & Path-loss exponent &   $2$\\ \hline
  WD Tx power  &   $2$ mW &  Carrier frequency &   $915$ MHz \\ \hline
  $DIFS$&   $50$ ms    & $PIFS$ &    $30$ ms\\ \hline
  $SIFS$&   $10$ ms    & $ERB$ &   $30$ ms\\ \hline
  $\sigma$&   $50$ ms    &  $ACK$        &   $20$ ms\\ \hline
  $T_{pl}$&   $420$ ms    &  $T_{et}$        &   $2.43$ s\\ \hline
  Transmit antenna gain &   $2.5$  & Receive antenna gain &   $2$\\ \hline
\end{tabular}
\end{center}
\label{stat}
\end{table}

\section{Simulation Results}
In this section, we use simulations to verify the analysis and evaluate the performance of the distributed scheduling protocol proposed. In all simulations, we use the Powercast TX91501-3W transmitter as the energy transmitter at the HAP and P2110 Powerharvester as the energy receiver at each WD with $\eta= 0.51$ energy harvesting efficiency.\footnote{Please see the detailed product specifications on the website of Powercast Co. (http://www.powercastco.com).} Unless otherwise stated, the simulation parameters are listed in Table I, which correspond to a typical outdoor sensor network. We consider two types of WDs, where type-I WDs are located around $5$ meters away from the HAP, while type-II WDs are located around $3.5$ meters away. From Table I, each WD consumes around $1$ mJ energy to transmit a payload. From (\ref{24}), type-I and type-II WDs harvest $1$ and $2$ units of energy, i.e., $e_n=1$ and $2$, respectively. Unless otherwise stated, we consider $18$ WDs with $12$ type-I WDs and $6$ type-II WDs. Besides, the battery capacity is set as $C=30$. Each point in the figures shown in this section is obtained from simulating the WPCN for $10^8$ time slots.

\begin{figure}
\centering
  \begin{center}
    \includegraphics[width=0.5\textwidth]{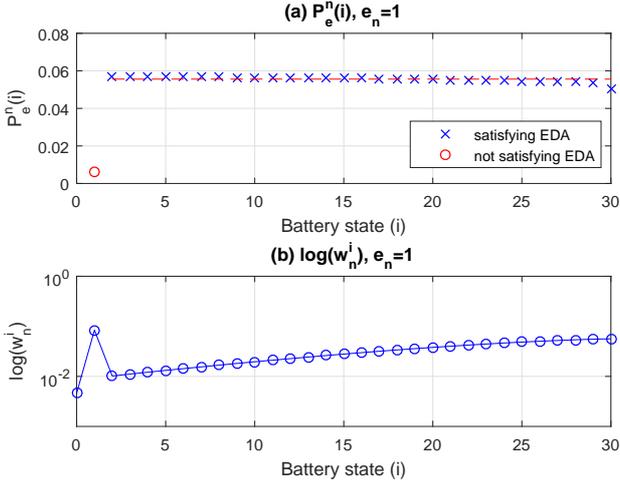}
  \end{center}
  \caption{Verification of the EDA assumption for WDs with $e_n=1$. Sub-figures (a) and (b) show $P_{e}^n$ and $w_n^i$ (in log-scale), respectively.}
  \label{105}
\end{figure}

\begin{figure}
\centering
  \begin{center}
    \includegraphics[width=0.5\textwidth]{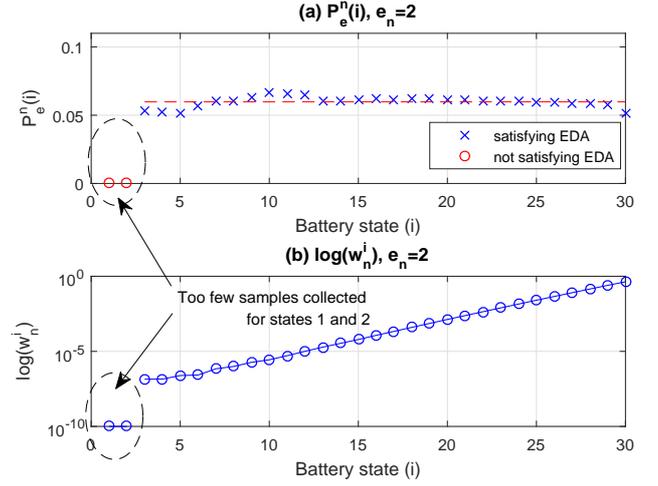}
  \end{center}
  \caption{Verification of the EDA assumption for WDs with $e_n=2$. Sub-figures (a) and (b) show $P_{e}^n$ and $w_n^i$ (in log-scale), respectively.}
  \label{107}
\end{figure}

We first verify in Figs.~\ref{105} and \ref{107} the proposed energy decoupling assumption (EDA). In particular, for each WD $n$, we calculate $P_e^n(i)$ by dividing the number of ET slots observed at each battery state $i$ and the number of occurrences of battery state $i$, $i=1,\cdots,C$. In Fig.~\ref{105}, we plot $P_e^n(i)$'s and the limiting probabilities of battery states ($w_n^i$'s) of type-I WDs. We can see that $p_e^n(i)$'s are approximately constant for $i=2,\cdots, C$, which matches the statement of EDA. The only exception is the $1$-st battery boundary state, where $p_e^n(1)$ is significantly lower than the other $p_e^n(i)$'s and $w_n^1$ is much higher than the other states. This is mainly due to its close connection with the $0$-th battery state, where a WD immediately enters the $1$-st battery state when it reaches the $0$-th battery state. We also plot in Fig.~\ref{107} the $P_e^n(i)$'s and $w_n^i$'s of type-II WDs. Interestingly, we can see that $p_e^n(i)$'s are approximately equal as long as sufficient samples are collected at energy state $i$, e.g., $3\leq i \leq 30$. For states $1$ and $2$, no sample or very few samples are collected due to the extremely low probabilities of the two states, thus the samples collected for the two states are ignored. From the above discussion, we can see that the proposed EDA approximately holds, which serves as the basis of our analysis.

\begin{figure}
\centering
  \begin{center}
    \includegraphics[width=0.5\textwidth]{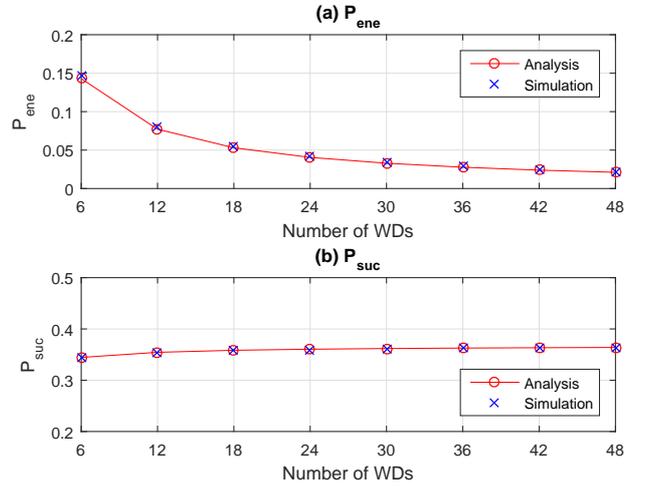}
  \end{center}
  \caption{Comparisons of analytical and simulation results. Sub-figures (a) and (b) show the probabilities of an energy transmission slot ($P_{ene}$) and successful transmission slot ($P_{suc}$), respectively.}
  \label{106}
\end{figure}

In Fig.~\ref{106}, we compare the throughput analysis in (\ref{21})-(\ref{23}) with simulations when the number of WDs changes from $N=6$ to $48$. Without loss of generality, we assume $\frac{1}{3}N$ WDs are type-I and the rest $\frac{2}{3}N$ WDs are type-II, and $p_t = \frac{1}{N}$. For brevity, we only present the results for the probability of a WET slot ($P_{ene}$ in (\ref{21})) and that of a successful transmission slot ($P_{suc}$ in (\ref{22})). We can see that the simulation and analysis match well, which validates our proposed analytical method. Besides, $P_{ene}$ decreases with the number of WDs, $N$. This is because, by setting $p_t = \frac{1}{N}$, the successful transmission probability keeps almost unchanged but each WD transmits less frequently. This reduces the overall energy consumption and the need for energy transfer. We can also infer from Fig. \ref{106} that the proposed distributed scheduling method can achieve stable throughput performance against the variation on the number of WDs, as long as the transmit probability $p_t$ is set proportionally to the number of WDs. In practice, the HAP can calculate $p_t$ by counting the number of associating WDs and broadcast its value to the WDs either periodically or when the number of associating WDs varies.

\begin{figure}
\centering
  \begin{center}
    \includegraphics[width=0.5\textwidth]{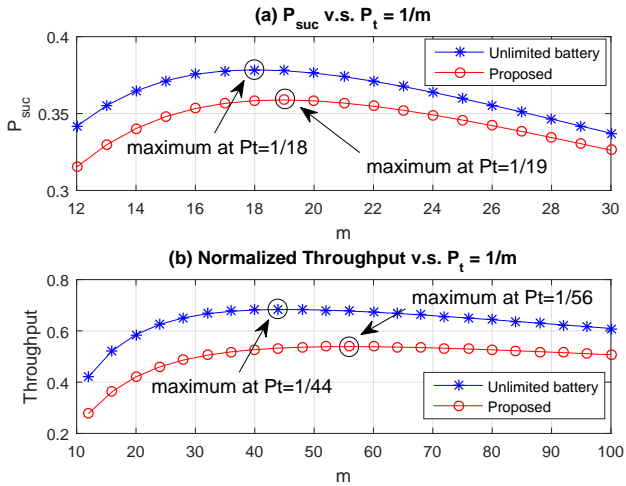}
  \end{center}
  \caption{Impact of $p_t = 1/m$ to the throughput performance. Sub-figures (a) and (b) show $P_{suc}$ and normalized throughput $\psi$, respectively.}
  \label{108}
\end{figure}

At last, we investigate the impact of data transmit probability ($p_t$) to the throughput performance. Here, we consider a performance benchmark with unlimited battery supply, i.e., no need of WET. This may correspond to the conventional $p$-persistent CSMA WLAN network without device energy constraint. Evidently, the benchmark method produces a performance upper bound of the energy-constrained scheme considered in this paper. All the points in Fig.~\ref{108} are calculated numerically based on the proposed analytical model. In Fig.~\ref{108}(a), we consider $N=18$ and compare $P_{suc}$ achieved by the two methods when $p_t = 1/m$, $m=12,\cdots,30$. We can see that the benchmark method achieves the maximum $P_{suc} \approx e^{-1}$ when $m=N=18$. The considered distributed scheduling, however, achieves the maximum $P_{suc}$ at a smaller $p_t$ when $m=19$. Besides, we also plot in Fig.~\ref{108}(b) the throughput performance comparison. Similarly, we can see that the maximum throughput of the proposed wireless-powered scheme is achieved at a smaller $p_t$ than that with unlimited energy supply, i.e., $p_t=\frac{1}{56}$ versus $\frac{1}{44}$. This is because a larger $p_t$ would induce high collision probability in both networks, but in WPCN, it also causes higher device energy consumption, and thus inducing more frequent WET and on average shorter airtime of WIT. Overall, the throughput of the WPCN is around $20\%$ lower than the case with unlimited energy supply, e.g., conventional WLAN. The performance loss is acceptable considering the additional overhead for WET in WPCN.

\section{Conclusions}
In this paper, we presented a fully distributed scheduling protocol for energy and information transmissions in RF-enabled WPCN. An energy queueing model was proposed to analyze the throughput performance, which leverages an interesting and novel energy decoupling property in the considered WPCN. Simulation results have verified our analysis and showed that the proposed distributed scheduling can achieve sustainable and efficient operation of WPCN.

\section*{Appendix\\Proof of the monotonic property of $f_n(x)$}
\emph{Proof}: Without loss of generality, we denote $w_n^i = \lambda_i w_n^0$ when $p_n^e=a$, and $w_n^i = \beta_i w_n^0$ when $p_n^e=b$, where $0<a\leq b<1$ and $i=1,\cdots,C$. Evidently, we can see from (\ref{14}) that
\begin{equation}
\lambda_1 = \frac{1}{(1-a)p_t} < \frac{1}{(1-b)p_t} = \beta_1.
\end{equation}
Similarly, by substituting $w_n^1$ into $w_n^2$ in (\ref{15}), we have
\begin{equation}
\begin{aligned}
\lambda_2 &= \frac{1}{(1-a)p_t} + \frac{a}{(1-a)p_t} \lambda_1 \\
&\leq \frac{1}{(1-b)p_t} + \frac{b}{(1-b)p_t} \lambda_1 \\
&\leq \frac{1}{(1-b)p_t} + \frac{b}{(1-b)p_t} \beta_1 = \beta_2.
\end{aligned}
\end{equation}
By repeatedly substituting $w_n^i$ into $w_n^{i+1}$ in either (\ref{15}) or (\ref{16}), we have
\begin{equation}
\lambda_i\leq \beta_i, \ i=1,\cdots,C.
\end{equation}
Therefore, we have
\begin{equation}
f_n(a) = \frac{1}{1+\sum_{i=1}^C \lambda_i} \geq \frac{1}{1+\sum_{i=1}^C \beta_i} = f_n(b),
\end{equation}
which leads to the proof of the desired result.  $\hfill\blacksquare$

%
%
%
%

\end{document}